\documentclass{article}
\usepackage{graphics}

\begin{document}

\newtheorem{thm1}{Theorem}
\newtheorem{lemma2}[thm1]{Lemma}
\newtheorem{lemma3}[thm1]{Lemma}
\newtheorem{lemma4}[thm1]{Lemma}
\newtheorem{cor5}[thm1]{Corollary}
\newtheorem{lemma6}[thm1]{Lemma}
\newtheorem{lemma7}[thm1]{Lemma}
\newtheorem{exempel}[thm1]{Example}
\newenvironment{expl}[1][{}]{\begin{exempel} {#1}\normalfont}{\end{exempel}}

\def\tr{ {\rm{Tr }}\,}
\def\pr{ {\rm{Pr }}}
\def\oti{{\otimes}}
\def\bra#1{{\langle #1 |  }}
\def\lb{ \left[ }
\def\rb{ \right]  }
\def\tilde{\widetilde}
\def\bar{\overline}
\def\*{\star} 

\def\({\left(}          \def\BL{\Bigr(}
\def\){\right)}         \def\BR{\Bigr)}
        \def\BBL{\lb}
        \def\BBR{\rb}
%
\newcommand{\qed}{\rule{7pt}{7pt}}
\def\E{{\mathbf{E} }}
\def\1{{\mathbf{1} }}

\def\bb{{\bar{b} }}
\def\ab{{\bar{a} }}
\def\zb{{\bar{z} }}
\def\zbar{{\bar{z} }}
\def\frac#1#2{{#1 \over #2}}
\def\inv#1{{1 \over #1}}
\def\half{{1 \over 2}}
\def\d{\partial}
\def\der#1{{\partial \over \partial #1}}
\def\dd#1#2{{\partial #1 \over \partial #2}}
\def\vev#1{\langle #1 \rangle}
\def\ket#1{ | #1 \rangle}
\def\rvac{\hbox{$\vert 0\rangle$}}
\def\lvac{\hbox{$\langle 0 \vert $}}
\def\2pi{\hbox{$2\pi i$}}
\def\e#1{{\rm e}^{^{\textstyle #1}}}
\def\grad#1{\,\nabla\!_{{#1}}\,}
\def\dsl{\raise.15ex\hbox{/}\kern-.57em\partial}
\def\Dsl{\,\raise.15ex\hbox{/}\mkern-.13.5mu D}
\def\b#1{\mathbf{#1}}
%
%
\def\th{\theta}         \def\Th{\Theta}
\def\ga{\gamma}         \def\Ga{\Gamma}
\def\be{\beta}
\def\al{\alpha}
\def\ep{\epsilon}
\def\vep{\varepsilon}
\def\la{\lambda}        \def\La{\Lambda}
\def\de{\delta}         \def\De{\Delta}
\def\om{\omega}         \def\Om{\Omega}
\def\sig{\sigma}        \def\Sig{\Sigma}
\def\vphi{\varphi}
%
%
\def\CA{{\cal A}}       \def\CB{{\cal B}}       \def\CC{{\cal C}}
\def\CD{{\cal D}}       \def\CE{{\cal E}}       \def\CF{{\cal F}}
\def\CG{{\cal G}}       \def\CH{{\cal H}}       \def\CI{{\cal J}}
\def\CJ{{\cal J}}       \def\CK{{\cal K}}       \def\CL{{\cal L}}

\def\CM{{\cal M}}       \def\CN{{\cal N}}       \def\CO{{\cal O}}
\def\CP{{\cal P}}       \def\CQ{{\cal Q}}       \def\CR{{\cal R}}
\def\CS{{\cal S}}       \def\CT{{\cal T}}       \def\CU{{\cal U}}
\def\CV{{\cal V}}       \def\CW{{\cal W}}       \def\CX{{\cal X}}
\def\CY{{\cal Y}}       \def\CZ{{\cal Z}}

\def\rvac{\hbox{$\vert 0\rangle$}}
\def\lvac{\hbox{$\langle 0 \vert $}}
\def\comm#1#2{ \BBL\ #1\ ,\ #2 \BBR }
\def\2pi{\hbox{$2\pi i$}}
\def\e#1{{\rm e}^{^{\textstyle #1}}}
\def\grad#1{\,\nabla\!_{{#1}}\,}
\def\dsl{\raise.15ex\hbox{/}\kern-.57em\partial}
\def\Dsl{\,\raise.15ex\hbox{/}\mkern-.13.5mu D}
\def\beq{\begin {equation}}
\def\eeq{\end {equation}}
\def\to{\rightarrow}

\title{Distilling common randomness from bipartite quantum states}

\author{I. Devetak\footnote{Electronic address: devetak@us.ibm.com} \\
\it{IBM T.J. Watson Research Center, Yorktown Heights, NY 10598, USA} \\
\\
 A. Winter\footnote{Electronic address: winter@cs.bris.ac.uk} \\
\it{Department of Computer Science, University of Bristol, Bristol BS8 1UB, U.K.}
}

  \date{\today} 
  \maketitle

\begin{abstract}
  The problem of converting noisy quantum correlations between two parties 
  into noiseless classical ones using a limited amount of one-way classical
  communication is addressed. A single-letter formula for the optimal 
  trade-off  between the extracted common randomness and classical communication 
  rate is obtained for the special case of classical-quantum correlations.
  \par
  The resulting curve is intimately related to the quantum compression
  with classical side information trade-off curve $Q^*(R)$ of Hayden, Jozsa
  and Winter.
  \par
  For a general initial state we obtain a similar result, with a single-letter
  formula, when we impose a tensor product restriction on the measurements
  performed by the sender; without this restriction the trade-off is given by
  the regularization of this function.
 \par
  Of particular interest is a quantity we call ``distillable common
  randomness'' of a state: the maximum overhead of the common randomness
  over the one-way classical communication if the latter is unbounded.
  It is an operational measure of (total) correlation in a quantum state.
  For classical-quantum correlations it is given by the Holevo mutual
  information of its associated ensemble, for pure states it is the
  entropy of entanglement. In general, it is given by an optimization
  problem over measurements and regularization; for the case of separable
  states we show that this can be single-letterized.

\end{abstract}

\section{Introduction}

\par 
Quantum, and hence also classical, information theory can be viewed as
a theory of inter-conversion between various resources. These
resources can be classical or quantum, static or dynamic,
noisy or noiseless. Based on  the number of spatially separated 
parties sharing a resource, it can be bipartite or multipartite; local (monopartite) 
resources are typically taken for granted. 
In what follows, we shall mainly be concerned with
bipartite resources. Let us introduce a notation in which
$c$ and $q$ stand for classical and quantum, respectively, 
curly and square brackets stand for noisy and noiseless, respectively, 
and arrows ($\rightarrow$) will distinguish dynamic resources
from static ones. The possible combinations are tabulated below.
Noisy dynamic resources are the four types of noisy channels, classified 
by the classical/quantum nature of the input/output. Beside the familiar
classical $\{ c \rightarrow c \}$  and quantum  $\{ q \rightarrow q \}$
channels, this category also includes preparation of quantum states from a given set 
(labeled by classical indices) $\{ c \rightarrow q \}$ and 
measurement of quantum states yielding classical outcomes $\{ q \rightarrow c \}$. 
Dynamic ``unit'' resources by definition require the input and output to be of 
the same nature, and they comprise of the noiseless bit $[c \rightarrow c ]$
and qubit $[q \rightarrow q]$ channel, but we additionally introduce
symbols for general (higher dimensional) perfect quantum and classical
channels: $(q \rightarrow q)$ and $(c \rightarrow c)$, respectively.

Noisy static resources, not having a directionality, can be one of three types:
classical $\{ c \, c \}$ , quantum $\{ q \, q \}$ and mixed classical-quantum
$\{ c \, q \}$. The first of these is embodied in a pair of 
correlated random variables $XY$, associated with the product set
$\CX \times \CY$ and a probability distribution $p(x,y) = \pr\{ X = x, Y = y\}$ 
defined on $\CX \times \CY $.  
The $\{ q \, q \}$ analogue is a bipartite quantum system $\CA \CB$,
associated with a product Hilbert space $\CH_\CA \otimes \CH_\CB$ 
and a density operator $\rho^{\CA \CB}$,
the ``quantum state'' of the system $\CA \CB$, defined on $\CH_\CA \otimes \CH_\CB$. 
A $\{ c \, q \}$ resource is a hybrid classical-quantum system
$X \CQ$, the state of which is now described by an \emph{ensemble} $\{  \rho_x, p(x) \}$,
with $p(x)$ defined on $\CX$ and the $\rho_x$ being density operators on
the Hilbert space $\CH_\CQ$ of $\CQ$. The state of the quantum system $\CQ$ is 
thus correlated with the classical index $X$.
A useful representation of $\{ c \, q \}$ resources, which
we refer to as the ``enlarged Hilbert space'' (EHS) representation,   
is obtained by embedding the random variable $X$ in some quantum system $\CA$.
Then our ensemble $\{ \rho_x, p(x) \}$ corresponds to the density 
operator 
\beq
\rho^{\CA \CQ} = \sum_x p(x) 
\ket{x}\bra{x}^{\CA} \otimes \rho_x^{\CQ},
\label{cq}
\eeq
where $\{\ket{x}: x \in \CX \}$ is an orthonormal basis for the Hilbert space 
$\CH_\CA$ of $\CA$. Thus $\{ c \, q \}$ resources may be viewed 
as a special case of $\{ q \, q \}$ ones.
Finally, we have noiseless static resources, which can be classical
$(c \, c)$ or quantum $(q \, q)$.
The classical resource is a pair of \emph{perfectly} correlated random variables,
which is to say that  $\CX = \CY$ and $p(x,y) = p(x) \delta(x,y)$
(without loss of generality).
We reserve the $[c \, c]$ notation for a \emph{unit} of common randomness (1 rbit),
a perfectly correlated pair of binary random variables with a full bit of entropy.
The quantum resource is a quantum system $\CA \CB$ in a pure entangled 
state $\ket{\psi}_{\CA \CB}$. Again, the $[ q \, q ]$ notation
denotes a unit of entanglement (1 ebit), a maximally entangled qubit
pair $\frac{1}{\sqrt{2}} (\ket{0}_\CA \ket{0}_\CB + \ket{1}_\CA \ket{1}_\CB )$.
Since $(c \, c)$ and $[c \, c]$, and 
$(q \, q)$ and $[q \, q]$ may be inter-converted in an asymptotically lossless
way and with an asymptotically vanishing rate of extra resources, for
most purposes it suffices to consider the unit resources only.
Note the clear hierarchy amongst unit resources: 
$$[ q \rightarrow q ]  \Longrightarrow  ( [ c \rightarrow c ] \,\, {\rm or} 
\,\, [ q \, q] )
\,\, \Longrightarrow [ c \, c]. $$
Any of the  conversions ($\Longrightarrow$) can be performed at a unit rate 
and no additional cost. On the other hand, $[ c \rightarrow c ]$  and 
$[ q \, q]$ are strictly ``orthogonal'': neither can be produced from the other.
\vspace{1mm}

\begin{center}
\begin{tabular}{|c|c|}
\hline
\multicolumn{2}{|c|}{Dynamic unit resources}  \\
\hline
\hline
$ [ c \rightarrow c ]$ & noiseless bit channel \\
\hline
$ [ q \rightarrow q ]$ & noiseless qubit channel \\
\hline
\end{tabular}
\end{center}

\begin{center}
\begin{tabular}{|c|c|}
\hline
\multicolumn{2}{|c|}{Noiseless dynamic resources}  \\
\hline
\hline
$ ( c \rightarrow c )$ & general noiseless channel --- w.l.o.g.~identity on some set\\
\hline
$ ( q \rightarrow q )$ & noiseless qubit channel --- w.l.o.g.~identity on some space\\
\hline
\end{tabular}
\end{center}

\begin{center}
\begin{tabular}{|c|c|}
\hline
 \multicolumn{2}{|c|}{Noisy dynamic resources}  \\
\hline
\hline
$\{ c \rightarrow c \}$ & noisy classical channel, given by a stochastic matrix $W$ \\
\hline
$\{ c \rightarrow q \}$ & quantum state preparation, given by quantum
alphabet $\{ \rho_x \}$  \\
\hline
$\{q \rightarrow c \}$ & generalized measurement, given by a POVM $(E_x)$  \\
\hline
$\{ q \rightarrow q \}$ & noisy quantum channel, given by CPTP map $\CN$ \\
\hline
\end{tabular}
\end{center}

\begin{center}
\begin{tabular}{|c|c|}
\hline
\multicolumn{2}{|c|}{Unit static resources}  \\
\hline
\hline
$[ c \, c ]$ & maximally correlated bits (1 rbit) \\
\hline
$[ q \, q ]$ & maximally entangled qubits (1 ebit) \\
\hline
\end{tabular}
\end{center}

\begin{center}
\begin{tabular}{|c|c|}
\hline
\multicolumn{2}{|c|}{Noiseless static resources}  \\
\hline
\hline
$( c \, c )$ & perfectly correlated random variables $X Y$ with distribution
$p(x,y) = p(x) \delta(x,y)$ \\
\hline
$( q \, q )$ & bipartite quantum system $\CA \CB$ in a pure state $\ket{\psi}_{\CA \CB}$ \\
\hline
\end{tabular}
\end{center}

\begin{center}
\begin{tabular}{|c|c|}
\hline
\multicolumn{2}{|c|}{Noisy static resources}  \\
\hline
\hline
$\{ c \, c \}$ & correlated random variables $X Y$ with joint distribution $p(x,y)$ \\
\hline
$\{ c \, q \}$ & classical-quantum system $X \CQ$ corresponding to an ensemble $\{\rho_x, p(x) \}$  \\
\hline
$\{ q \, q \}$ & bipartite quantum system $\CA \CB$ in a general quantum state 
$\rho^{\CA \CB}$ \\
\hline
\end{tabular}
\end{center}

The generality of this classification is illustrated in
the table below, where the  resource inter-conversion task is identified 
for a number of examples from the literature.
To interpret these ``chemical reaction formulas'', there is but
rule to obey: if non-unit resources appear on the right, then all non-unit
(dynamical) resources are meant to be fed from some fixed source.
For example, $(c \rightarrow c)$ in the output of a transformation
symbolizes the noiseless transmission of an implicit classical information
source, and likewise $(q \rightarrow q)$
the noiseless transmission of an implicit quantum information source

\begin{center}
\begin{tabular}{|l|l|}
\hline
 \multicolumn{2}{|c|}{Some known problems in classical and quantum information theory} \\
\hline
\hline
$[ c \rightarrow c ] \Longrightarrow ( c \rightarrow c )$ &
Shannon compression \cite{shannon} \\
\hline
$[ q \rightarrow q ] \Longrightarrow ( q \rightarrow q )$ &
Schumacher compression \cite{nono}\\
\hline
$( q \, q ) \Longrightarrow [ q \, q ] $ & Entanglement concentration \cite{bbps} \\
\hline
$[ q \, q ] + [c \leftrightarrow c] \Longrightarrow ( q \, q ) $ & 
Entanglement dilution \cite{bbps, lo, koren}\\
\hline
$[ q \, q ] + [c \leftrightarrow c]  \Longrightarrow \{ q \, q \} $ & 
Entanglement cost, entanglement of \\ & purification \cite{bdsw, hht, thld} \\
\hline
$\{ q \, q \} + [c \leftrightarrow c]  \Longrightarrow [ q \, q ] $ & 
Entanglement distillation \cite{bdsw,q}\\
\hline
$\{ c \, c \} + [c \leftrightarrow c]
 \Longrightarrow [ c \, c ] $ & 
Classical common randomness capacity \cite{ac1, ac2} \\
\hline
$\{ c \, q \} + [c \rightarrow c]  \Longrightarrow [ c \, c ] $ & 
{\bf present paper} \\
\hline
$\{ q \, q \} + [c \rightarrow c]  \Longrightarrow [ c \, c ] $ & 
{\bf present paper} \\
\hline
$\{ c \rightarrow c \} \Longrightarrow [c \rightarrow c] $ & 
Shannon's channel coding theorem \cite{shannon} \\
\hline
$\{ c \rightarrow q \} \Longrightarrow [c \rightarrow c] $ & 
HSW theorem (fixed alphabet) \cite{hsw} \\
\hline
$\{ q \rightarrow q \} \Longrightarrow [c \rightarrow c] $ & 
HSW theorem (fixed channel) \cite{hsw}\\
\hline
$\{ q \rightarrow q \} \Longrightarrow [q \rightarrow q] $ & 
Quantum channel coding theorem \cite{q}\\
\hline
$[ c \rightarrow c ]  + [ q \, q ] \Longrightarrow [q \rightarrow q] $ & 
Quantum teleportation \cite{tele}\\
\hline
$[ q \rightarrow q ]  + [ q \, q ] \Longrightarrow [c \rightarrow c] $ & 
Quantum super-dense coding \cite{dense}\\
\hline
$\{ q \rightarrow q \}  + [ q \, q ] \Longrightarrow [c \rightarrow c] $ & 
Entanglement assisted classical capacity \cite{eac}\\
\hline
$\{ q \rightarrow q \}  + [ q \, q ] \Longrightarrow [q \rightarrow q] $ & 
Entanglement assisted quantum capacity \cite{eac} \\
\hline
$[ c \rightarrow c ]  + [ c \, c ] \Longrightarrow \{ c \rightarrow c \} $ & 
Classical reverse Shannon  theorem \cite{eac} \\
\hline
$[ c \rightarrow c ]  + [ q \, q ] \Longrightarrow \{ q \rightarrow q \} $ & 
Quantum reverse Shannon theorem \cite{qrst} \\
\hline
$\{ q \rightarrow c \}  + [ c \, c ] \Longrightarrow \{ q \rightarrow c \} $ & 
Winter's POVM compression theorem \cite{winter} \\
\hline
$[ c \rightarrow c ]  + [ q \, q ] \Longrightarrow \{ c \rightarrow q \} $ & 
Remote state preparation \cite{rsp, devberger} \\
\hline
$[ c \rightarrow c ] + [ q \rightarrow q ] \Longrightarrow \{ c \rightarrow q \}$ & 
Quantum-classical trade-off in quantum \\ & data compression \cite{hjw} \\
\hline
$\{ c \rightarrow q \} + [ c \rightarrow c ] \Longrightarrow ( c \rightarrow c )$ & 
Classical compression with quantum \\ & side information \cite{pqsw} \\
\hline
\end{tabular}
\end{center}

The present paper addresses the static ``distillation'' 
(noisy $\Longrightarrow$ noiseless) task of converting
noisy quantum correlations $\{q \, q \}$, i.e. bipartite quantum states,
into noiseless classical ones $[c \, c]$, i.e. common randomness (CR).
Many information theoretical problems are motivated by simple 
intuitive questions. For instance, Shannon's channel coding theorem 
\cite{shannon} 
quantifies the ability of a channel to send information. Similarly,
our problem stems from the desire to quantify the classical correlations
present in a bipartite quantum state. A recent paper by Henderson and
Vedral \cite{hv} poses this very question, and introduces several plausible 
measures. However, the     
ultimate criterion for accepting something as an information measure is 
whether it appears in the solution to an appropriate asymptotic 
information processing task; in other words, whether is has an \emph{operational}
meaning. It is this operational approach that is pursued here.
 
 The structure of our conversion problem is akin to two other 
static distillation problems:  $\{q \, q \} \Longrightarrow [q \, q]$
and  $\{c \, c \} \Longrightarrow [c \, c]$.
The former goes under the name of
``entanglement distillation'': producing maximally 
entangled qubit states from a large number of copies of  $\rho^{\CA \CB}$
with the help  of unlimited one-way or two-way classical communication 
\cite{bdsw}.
Allowing free classical communication in these problems is legitimate since,
as already noted, entanglement and classical communication are orthogonal resources.  
The $\{c \, c \} \Longrightarrow [c \, c]$ problem is one of creating 
CR from general correlated random variables, which is known to be impossible 
without additional classical communication. 
Now allowing free communication is
inappropriate, since it could be used to create unlimited CR.
There are at least two scenarios that do make sense, however,
and have been studied by Ahlswede and Csisz\'{a}r
in \cite{ac1} and \cite{ac2}, respectively.
In the first, one makes a distinction 
between the distilled key, which is required to be secret, 
and the classical communication which is public.
The second scenario involves limiting the amount of classical communication
to a one-way rate of $R$ bits per input state and asking about 
the maximal CR  generated in this way (see \cite{ac2}
for further generalizations). 
One can thus think of the classical communication as a quasi-catalyst
that enables distillation of a part of the noisy correlations, 
while itself becoming CR; it is not a genuine catalyst 
because the original dynamic resource is more valuable
than the static one.
We find that these classical results 
generalize rather well to our information processing task. 
The analogue of the first scenario \cite{ac1} has been
treated in an unpublished paper by Winter and Wilmink \cite{ww}. 
In this paper we generalize \cite{ac2}.
As a corollary we give one (of possibly many) operationally motivated answers to the 
question ``How much classical correlation is there in a bipartite 
quantum state?''. 
 
Alice and Bob share $n$ copies (in classical jargon: an $n$ letter word)
of a bipartite quantum state $\rho^{\CA \CB}$ . 
Alice is allowed $nR$ bits of classical 
communication to Bob. The question is: how much CR can 
they generate under these conditions? More precisely, Alice is allowed to perform
some measurement on her part of $(\rho^{\CA \CB})^{\otimes n}$, producing the 
outcome random variable $X^{(n)}$ defined on some set $\CX^{(n)}$. Next, she sends 
Bob $f(X^{(n)})$, where $f: \CX^{(n)} \rightarrow \{ 1,2, \dots, 2^{nR} \}$.
 The \emph{rate} $R$ signifies the number of bits per letter needed 
to convey this information.
Conditioned on the value of $f(X^{(n)})$, Bob performs an appropriate measurement 
with outcome random variable $Y^{(n)}$. We say that a pair of random variables 
$(K, L)$, both taking values in some set $\CK$, is \emph{permissible} if 
\begin{eqnarray}
K\! & = & K(X^{(n)}) \nonumber\\  
L & = & L(Y^{(n)}, f(X^{(n)})). \nonumber  
\end{eqnarray}  
A permissible
pair $(K, L)$ represents \emph{$\epsilon$-common randomness} if 
\beq
{\rm Pr} ( K \neq L) \leq \epsilon.
\label{komon}
\eeq
In addition we require the technical condition 
that $K$ and $L$ are in the same set satisfying
\beq
|\CK| \leq 2^{c' n}
\label{kardi}
\eeq
for some constant $c'$. Thus, strictly speaking, our CR
is of the $(c \, c)$ type, but it can easily be converted 
to $[c \, c]$ CR via local processing (intuitively, we would like to
say ``Shannon data compression'', only that the randomness thus obtained
is not uniformly distributed but ``almost uniformly'' in the sense of
the AEP~\cite{coverthomas}).
A CR-rate pair $(C, R)$ of common randomness $C$ and classical
side communication $R$ is called \emph{achievable} if for all $\epsilon, \delta > 0$
and all sufficiently large $n$ there exists a permissible pair $(K,L)$
satisfying (\ref{komon}) and (\ref{kardi}), such that 
$$
 \frac{1}{n}H(K) \geq C - \delta.
$$
We define the CR-rate function $C(R)$ to be
 $$ C(R) = \sup \{C: (C, R) \,\, {\rm {is}} \,\, {\rm{ achievable}}  \}.$$
One may also formulate the $C(R)$ problem for Alice and Bob
sharing some classical-quantum resource $X \CQ$ rather than the
fully quantum $\CA \CB$. In this case Alice's
measurement is omitted  since she already has the classical random variable
$X^{(n)} = X^n$. In the original classical problem \cite{ac2}
Alice and Bob share the classical resource $XY$.
There Bob's measurement is also omitted, since he already has the 
random variable $Y^{(n)} = Y^n$. 
Finally, we introduce the \emph{distillable} CR as 
\beq
D(R) = C(R) - R,
\eeq 
the amount of CR generated in excess of the invested classical 
communication rate. 
This suggests $D(\infty)$ as a natural asymmetric measure of the total 
classical correlation in the state.
As we shall see, the above turns out to be equivalent to the
asymptotic (``regularized'') version of $C_{\CA} (\rho^{\CA \CB})$, 
as defined in \cite{hv}.

The paper is organized as follows. First we consider the 
special case of $\{ c \, q \}$  resources 
for which evaluating $C(R)$ reduces to a single-letter optimization problem. 
Then we consider the $\{ q \, q \}$
case which builds on it rather like the fixed channel
Holevo-Schumacher-Westmoreland (HSW) theorem builds on the fixed alphabet
version.

\section{Classical-quantum correlations}

In this section we shall assume that Alice and Bob
share $n$ copies of some $\{c, q \}$ resource $X \CQ$, 
defined by the ensemble $\CE = \{ \rho_x, p(x) \}$ or, equivalently, equation ($\ref{cq}$).
Alice knows the random variable $X$ and Bob possesses the $d$-dimensional quantum system $\CQ$.
In what follows we shall make use of the EHS representation
to define various information theoretical
quantities for classical-quantum systems. The von Neumann entropy of a quantum 
system $\CA$ with density operator 
$\rho^\CA$ is defined as $H(\CA) = - \tr \rho^\CA \log \rho^\CA$.
For a bipartite quantum system $\CA \CB$ define formally the quantities
\emph{conditional von Neumann entropy}
$$
H(\CB| \CA) = H(\CA \CB) - H(\CA),
$$
and \emph{quantum mutual information} (introduced earlier
as ``correlation entropy'' by Stratonovich )
$$
I(\CA; \CB) = H(\CA) + H(\CB) - H(\CA \CB) = H(\CB) - H(\CB| \CA).
$$
For general states of $\CA \CB$ we introduce these quantities without
implying an operational meaning for them.
(Though the quantum mutual information appears in the entanglement
assisted capacity of a quantum channel \cite{eac}, and the negative of
the conditional entropy, known as the coherent information
appears in the quantum channel capacity \cite{bdsw,q}.)

Introducing these quantities in formal analogy has the virtue of
allowing us to use the familiar identities and many of the inequalities
known for classical entropy. This to us seems better than claim any particular
operational connection (which, by all we known about quantum information
today, cannot be unique anyway).

Subadditivity of von Neumann entropy 
implies $I(\CA; \CB) \geq 0$.
For a tripartite quantum system $\CA \CB \CC$ define the quantum conditional
mutual information 
$$
I(\CA; \CB| \CC) = H(\CA| \CC) + H(\CB| \CC) - H(\CA \CB| \CC) 
                 = H(\CA\CC) + H(\CB\CC) - H(\CA\CB\CC) - H(\CC).
$$
Strong subadditivity of von Neumann entropy implies 
$I(\CA; \CB| \CC) \geq 0$.
A commonly used identity is the chain rule
$$
I(\CA; \CB \CC) = I(\CA; \CB) + I(\CA; \CC | \CB).
$$
Notice that for classical-quantum correlations ($\ref{cq}$)
the von Neumann entropy $H(\CA)$ is just the
Shannon entropy $H(X)$ of $X$. 
We define  the  mutual information of a classical-quantum system 
$X \CQ$ as $I(X; \CQ) = I(\CA; \CQ)$. Notice that this is
no other than the Holevo information of the ensemble $\CE$
$$\chi(\CE) = H\left( \sum_x p(x) \rho_x \right) - \sum_x p(x) H(\rho_x).$$ 
(Even though Gordon and Levitin have written down this expression much
earlier --- see \cite{holevo:coding} for historical
references ---, we feel that the honour should be with Holevo
for his proof of the information bound named duly after him \cite{holevo}.)

Using the EHS
representation for some tripartite  classical-quantum system
$U X \CQ$, strong subadditivity \cite{lieb} gives inequalities such as
$I(U;X|\CQ) \geq 0$ or $I(U; \CQ | X) \geq 0$
and the chain rule implies, e.g.,
$$
I(U;X \CQ) =  I(U;\CQ) + I(U;X |\CQ). 
$$
We shall take such formulae for granted throughout the paper. 

An important classical concept is that of a Markov chain 
of random variables $U \rightarrow X \rightarrow Y$ whose probabilities 
obey $\pr\{Y = y | X = x, U = u\} = \pr\{Y = y | X = x\}$, which is to say
that $Y$ depends on $U$ only through $X$. Analogously we may define a
classical-quantum
Markov chain $U \rightarrow X \rightarrow \CQ$ associated with an ensemble
$\{  \rho_{ux}, p(u,x) \}$ for which $\rho_{ux} = \rho_x$. Such an object 
typically comes about by augmenting the system $X \CQ$ by the
random variable $U$ (classically) correlated with $X$ via a conditional
distribution $Q(u|x) = \pr\{U = u | X = x\}$. In the EHS representation
this corresponds to the state
\beq
\rho^{\CZ \CA \CQ} = \sum_x p(x) \sum_u Q(u|x) 
\ket{u}\bra{u}^{\CZ}  \otimes
\ket{x}\bra{x}^{\CA} \otimes \rho_x^{\CQ}.
\label{drei}
\eeq
We are now ready to state our main result.
\begin{thm1}[CR-rate theorem for classical-quantum correlations] \label{t1}
\beq
 C(R) = {C}^*(R) = R + D^*(R),
 \label{main0}
\eeq
where 
\beq
 D^*(R)= \sup_{U|X} \{ I(U;\CQ)\, | \, I(U;X) - I(U;\CQ) \leq R \}. 
 \label{main}
\eeq
The supremum is to be understood as one over all conditional probability 
distributions $p(u|x)$ for the random variable $U$ conditioned on $X$,
with finite range $\CU$.
We may in fact restrict to the case $|\CU| \leq |\CX| + 1$, which in particular
implies that the $\sup$ is actually a $\max$.
\end{thm1}

The proof of the theorem is divided into two parts: show that $C^*(R)$ is an 
upper bound (commonly called the
``converse'' theorem) for $C(R)$, and then providing a direct coding scheme
demonstrating its achievability. We start with a couple of lemmas.

\vspace{3mm}

\begin{lemma6} \label{t6}
$D^*(R)$, and hence ${C}^*(R)$,
is monotonically increasing and concave; the latter meaning that
for $R_1,R_2\geq 0$ and
  $0\leq\lambda\leq 1$,
$$
\lambda {D}^*(R_1)+(1-\lambda){D}^*(R_2)
                    \leq {D}^*\bigl( \lambda R_1 +(1-\lambda)R_2 \bigr).
$$
\end{lemma6}
\noindent  $\mathbf{Proof }  $ \space \space 
The monotonicity of $D^*(R)$ is obvious from its definition.
To prove concavity, choose $U_1$, $U_2$ feasible for $R_1$, $R_2$, 
respectively: in particular,
\begin{eqnarray}
I(U_1;X)- I(U_1;\CQ)  & \leq & R_1, \nonumber\\
I(U_2;X)-  I(U_2;\CQ) & \leq & R_2.  \nonumber
\end{eqnarray}
Then, introducing the new random variable
$$
U =  \left\{ \begin{array}{ll}
(1,U_1) & \rm{with \,\, probability \,\,}\lambda, \\
(2,U_2) & \rm{with \,\, probability \,\,}1-\lambda,
\end{array} \right.
$$
we have
\begin{eqnarray}
\lambda I(U_1;X)+(1-\lambda)I(U_2;X) & =  & I(U;X), \nonumber\\
\lambda I(U_1;\CQ)+(1-\lambda)I(U_2;\CQ) & = & I(U;\CQ).  \nonumber
\end{eqnarray}
Thus
$$
\lambda I(U_1;\CQ)+(1-\lambda)I(U_2;\CQ) = I(U;\CQ) \leq D^*(R),
$$
the last step from
$$
I(U;X) - I(U;\CQ) \leq \lambda R_1 +(1-\lambda) R_2  \leq R.
$$
\qed
\vspace{3mm}

Consider the $n$ copy classical-quantum system $X^n \CQ^n = X_1 \CQ_1 X_2 \CQ_2 
\dots X_n \CQ_n$, in the state given by the $n$th tensor power of 
the ensemble $\{ \rho_x, p(x) \}$. Define now
\beq
 {D}_n^*(R)= \max_{U|X^n} \left\{ \frac{1}{n} I(U;\CQ^n)\, \big| \,\frac{1}{n}
                            \left( I(U;X^n) - I(U;\CQ^n) \right) \leq R \right\}. 
\label{nkopi}
\eeq
It turns out that this expression may be ``single-letterized'':
$$D_n^*(R) = D^*(R).$$
We prove slightly more by showing the following lemma, which implies the above equality
by iterative application and then using concavity of $D^*$ in $R$ (lemma \ref{t6}):
\begin{lemma7} \label{t7}
For two ensembles $\CE_1=\{\rho_x,p(x)\}$ ($x\in{\cal X}_1$)
and $\CE_2=\{\sigma_{x'},p'(x')\}$ ($x'\in{\cal X}_2$), denote their respective
$D^*$ functions $D^*(\CE_1,R)$ and $D^*(\CE_2,R)$. Then
$$D^*(\CE_1\otimes\CE_2,R) = \max \bigl\{ D^*(\CE_1,R_1)+D^*(\CE_2,R_2)
                                                   \,|\, R_1+R_2=R \bigr\}.$$
\end{lemma7}

\noindent  {\bf Proof}   \space \space
Let $\CE_1$ and $\CE_2$ correspond to the classical-quantum systems
$X_1 \CQ_1$ and $X_2 \CQ_2$, respectively. As before, we augment
the joint system by the random variable $U$ via the conditional
distribution $Q(u|x x')$, so that $U X_1 X_2 \CQ_1 \CQ_2$ obeys
the Markov property $U \rightarrow  X_1X_2 \rightarrow \CQ_1\CQ_2$.
In the EHS representation we have
$$\rho^{\CZ\CA_1\CA_2\CQ_1\CQ_2} = \sum_{u,x,x'} p(x)p'(x')Q(u|xx')
                                      \ket{u}\bra{u}^{\CZ}\otimes
                                      \ket{x}\bra{x}^{\CA_1}\otimes\ket{x'}\bra{x'}^{\CA_2}\otimes
                                      \rho_x^{\CQ_1}\otimes\sigma_{x'}^{\CQ_2}.$$
By definition, $D^*(\CE_1\otimes\CE_2,R)$ equals $I(U;\CQ_1\CQ_2)$ maximized
over all variables $U$ such that $I(U;X_1X_2) - I(U;\CQ_1\CQ_2) \leq R$.
\par
Now the inequality ``$\geq$'' in the lemma is clear: for we could
choose $U_1$ optimal for $\CE_1$ and $R_1$ and $U_2$ optimal for $\CE_2$ and $R_2$,
and form $U=U_1U_2$. By elementary operations with the definition of $D^*$
we see that $D^*(\CE_1,R_1)+D^*(\CE_2,R_2)$ is achieved.
\par
For the reverse inequality, let $U$ be any variable such that
$I(U;X_1X_2) - I(U;\CQ_1\CQ_2) \leq R$. First note that the Markov 
property $U \rightarrow  X_1X_2 \rightarrow \CQ_1\CQ_2$
implies  $I(U;X_1X_2) = I(U; X_1\CQ_1 X_2\CQ_2)$, which can
easily be verified in the EHS representation. 
Intuitively, possessing $\CQ_1\CQ_2$ in addition to knowing $X_1X_2$
conveys no extra information about $U$. Hence, by the chain rule,
$$I(U;X_1X_2) - I(U;\CQ_1\CQ_2) = I(U;X_1X_2|\CQ_1\CQ_2).$$
Now, using the chain rule and once more the fact that the content of
$\CQ_1$ is a function of $X_1$, we estimate
\begin{eqnarray}
  R & \geq & I(U;X_1X_2|\CQ_1\CQ_2) \nonumber\\
    & =    & I(U;X_1|\CQ_1\CQ_2) + I(U;X_2|\CQ_1\CQ_2 X_1)  \nonumber\\
    & =    & I(U;X_1|\CQ_1\CQ_2) + I(U;X_2|\CQ_2 X_1).  \nonumber\\
    & \geq & I(U;X_1|\CQ_1) + I(U;X_2|\CQ_2 X_1). \nonumber
\end{eqnarray}
Here the inequality of the last line is obtained by the following reasoning:
\begin{eqnarray}
  I(U;X_1|\CQ_1\CQ_2) & =    & I(U\CQ_2;X_1|\CQ_1) - I(X_1;\CQ_2|\CQ_1) \nonumber\\
                      & \geq & I(U;X_1|\CQ_1)      - 0, \nonumber
\end{eqnarray}
using strong subadditivity and the fact that $X_1\CQ_1-X_2\CQ_2$ is in a product
state.
\par
Hence there are $R_1$ and $R_2$ summing to $R$ for which
\begin{eqnarray}
  I(U;X_1)-I(U;\CQ_1)         = I(U;X_1|\CQ_1)     & \leq & R_1,   \label{R1} \\
  I(U;X_2|X_1)-I(U;\CQ_2|X_1) = I(U;X_2|\CQ_2 X_1) & \leq & R_2.   \label{R2}
\end{eqnarray}
On the other hand,
\begin{eqnarray}
  I(U;\CQ_1\CQ_2) & =    & I(U;\CQ_1) + I(U;\CQ_2|\CQ_1)                 \nonumber\\
                  & =    & I(U;\CQ_1) + I(U\CQ_1;\CQ_2) - I(\CQ_1;\CQ_2) \nonumber\\
                  & \leq & I(U;\CQ_1) + I(UX_1;\CQ_2)                    \nonumber\\
                  & =    & I(U;\CQ_1) + I(X_1;\CQ_2) + I(U;\CQ_2|X_1)    \nonumber\\
                  & =    & I(U;\CQ_1) + I(U;\CQ_2|X_1), \label{I1I2}
\end{eqnarray}
using the chain rule repeatedly; the inequality comes from the quantum analogue of the familiar
\emph{data processing inequality}~\cite{ahlswede:loeber}, another consequence
of the content of $\CQ_1$ being a function of $X_1$.
With (\ref{R1}) and by definition of $D^*$, $I(U;\CQ_1) \leq D^*(\CE_1,R_1)$.
But also, with (\ref{R2}), $I(U;\CQ_2|X_1) \leq D^*(\CE_2,R_2)$, observing that
the conditional mutual information in (\ref{I1I2}) as well as in (\ref{R2})
are probability averages over unconditional mutual informations, and invoking
the concavity of $D^*$ (lemma \ref{t6}).
\par
Hence,
$$I(U;\CQ_1\CQ_2) \leq D^*(\CE_1,R_1)+D^*(\CE_2,R_2),$$
and since $U$ was arbitrary, we are done.
\qed

\vspace{3mm}

\noindent  {\bf Proof of Theorem 1 (converse) }   \space \space 
For a given blocklength $n$, measurement on Bob's side will turn the classical-quantum
correlations into classical ones, and $\CQ^n$ gets replaced by the measurement 
outcome random variable $Y^{(n)}$. Now we can apply the 
classical converse \cite{ac2} to the classical random variable pair 
$(X^n, Y^{(n)})$
$$
C(R) \leq R + \max_{U|X^{n}} \left\{ \frac{1}{n} I(U;Y^{(n)}) \, | \, 
I(U;X^n) - I(U;Y^{(n)}) \leq n R \right\}.
$$
By the  the Holevo inequality \cite{holevo} 
$$I(U;Y^{(n)}) \leq  I(U;{\CQ^n}),$$
this can be further
bounded by $C^*_n(R)$ which is, by lemma \ref{t7}, equal to $C^*(R)$.
To complete the proof, we need to show that the supremum
in (\ref{main}) can be restricted to a set $\CU$ of cardinality
$|\CU| \leq |\CX| + 1$. This is a standard consequence of
Caratheodory's theorem, and the proof runs in exactly the same
way as that in, e.g., \cite{hjw}.
\qed
\vspace{2mm}

\noindent We shall need some auxiliary results before we embark on
proving the achievability of $C^*(R)$.

\begin{lemma2} \label{t2}
The $(C,R)$ pair $(H(X), H(X|\CQ))$ is achievable when
Alice and Bob share the classical-quantum system $X \CQ$.
\end{lemma2}
\noindent  $\mathbf{Proof }  $ \space \space This follows from the 
classical-quantum Slepian-Wolf result \cite{pqsw} which states 
that, for any  $\epsilon, \delta  > 0$ and sufficiently large $n$, 
the classical communication rate from Alice to Bob sufficient for 
Bob to reproduce $X^n$ with error probability $\leq \epsilon$  
is $H(X|\CQ) + \delta$. \qed

\vspace{2mm}

\noindent $\mathbf{Remark }  $ \space \space  
Lemma \ref{t2} already
yields the value of $$D(\infty) = D(H(X|\CQ)) = H(X) - H(X|\CQ) =
I(X; \CQ)$$
for the classical-quantum system $X \CQ$. This justifies our
interpretation of $D(\infty)$ as the amount of classical correlation
in $X \CQ$.

\begin{lemma3} \label{t3}
Let $\sigma$ be a state in a $D$-dimensional Hilbert space.
Then  $\tr(\sigma B) = 1 - \epsilon$ for some operator 
$0 \leq B \leq \1$ implies
\beq
  H(\sigma) \leq 1 + \epsilon \log D + (1 - \epsilon) \log(\tr B + 1)
\label{daga}
\eeq
\end{lemma3}

\noindent  $\mathbf{Proof }  $ \space \space 
Diagonalize $\sigma$ as $\sigma = \sum_{j = 1}^D p_j \ket{j} \bra{j}$
with $p_1 \leq  p_2 \dots \leq p_D$ and define $b_j = \bra{j} B \ket{j}$,
so that
\beq
\sum_j p_j b_j = 1 - \epsilon
\label{nida}
\eeq 
and $\tr B = \sum_j b_j$.
Further define the random variable $J$ with ${\rm{Pr}}\{J = j\} = p_j$, 
for which $H(\sigma) = H(J)$. Consider the vector $\tilde{b}^D$ which minimizes
$\sum_j b_j$ subject to  constraints (\ref{nida})
and $0 \leq b_j \leq 1$. This is a trivial linear programming problem,
solved at the boundary of the allowed region for the $b_j$.
It is easily verified that the solution is given by 
\begin{eqnarray}
 & \tilde{b}_1 = \dots  = \tilde{b}_{k-1} = 0 , \nonumber\\
& 0 \leq \tilde{b}_k \leq 1, \nonumber\\
& \tilde{b}_{k+1} = \dots  = \tilde{b}_D = 1,\nonumber
\end{eqnarray}
for some $1 \leq k \leq D$ for which (\ref{nida}) is
satisfied. Note that $$D - k \leq \sum_j \tilde{b}_j \leq \tr B$$ and 
$\sum_{j = 1}^{k - 1} p_j \leq \epsilon$. Define the indicator random
variable $I(J)$
$$
I(J) =  \left\{ \begin{array}{ll}
1 & J \geq k, \\
0 & \rm{otherwise.}
\end{array} \right.
$$
We then have
\begin{eqnarray}
H(J) & = &  H(I) + H(J|I)  \nonumber\\
& \leq & 1 + {\rm{Pr}}\{J = 0\} \log D +  
{\rm{Pr}}\{J = 1\}  \log (D + 1 - k) \nonumber\\
& \leq & 1 + \epsilon \log D + (1 - \epsilon) \log (\tr B + 1),\nonumber
\end{eqnarray}
which proves the lemma.
\qed 

\vspace{2mm}

In order to understand the next two results, some 
background on typical sets $\CT^n_{U,\delta}$, conditionally typical sets 
$\CT^n_{X|U,\delta}(u^n)$, typical subspaces $\Pi^n_{\CQ, \delta}$ and
conditionally typical subspaces $\Pi^n_{\CQ|U, \delta}(u^n)$ is needed 
\cite{ck, nono, winter}. This is provided in the Appendix.

\begin{lemma4} \label{t4}
For every $\epsilon, \delta > 0$ and set $\CE  \subset \CX^n$ with 
${\rm{Pr}}\{X^n \in \CE\} \geq \epsilon$, there exists a subset 
$\CF \subset \CE$ and a sequence $u^n \in \CT^n_{U, \delta}$ such that
\beq
\CF  \subset \CT^n_{X|U, \delta}(u^n), \,\,\,\,\,\,\,\,\,
 \left | \frac{1}{n} \log |\CF|  - H(X|U) \right | \leq \delta,
\label{druga}
\eeq
whenever $n \geq n_1(|\CU|,|\CX|,\epsilon, \delta)$.
In addition, whenever  $n \geq n_2(|\CU|,|\CX|,d,\epsilon, \delta)$,
\beq
\frac{1}{n} H(\CQ^n| X^n \in \CF) \leq H(\CQ|U) + \delta.
\label{treca}
\eeq
\end{lemma4}

\noindent  $\mathbf{Proof }  $ \space \space 
Clearly, it suffices to prove the claim for some sufficiently small 
$\epsilon$.
The first claim
(\ref{druga}) is a purely classical result
and corresponds to lemma 3.3.3  of Csisz\'{a}r and K\"orner \cite{ck}.
Thus it remains to demonstrate (\ref{treca}).
We shall need the following facts from the Appendix.
For sufficiently large
$n \geq n_0(|\CU|,|\CX|,d,\delta',\epsilon)$,
for $x^n \in \CT^n_{X|U,\delta'}(u^n)$ and $u^n \in \CT^n_{U, \delta'}$:
\beq
\tr(\rho_{u^n x^n} \Pi^n_{\CQ | U,(|\CX| + 1)\delta'}(u^n)) \geq 1 - \epsilon,
\label{babel}
\eeq
and
\beq
\tr \Pi^n_{\CQ | U,(|\CX| + 1)\delta'}(u^n) \leq 2^{n H(\CQ|U) + (2 + |\CX|) c \delta'}.
\label{mabel}
\eeq
Since $\rho_{u^n x^n} = \rho_{x^n}$, it follows from the linearity of
trace and (\ref{babel}) that
$$
\tr(\rho_{\CF} \Pi^n_{\CQ | U,(|\CX| + 1)\delta'}(u^n)) \geq 1 - \epsilon,
$$ 
where 
$$
\rho_{\CF} = \sum_{x^n} {\rm{Pr}}\{X^n = x^n|X^n \in \CF\} \rho_{x^n}. 
$$ 
Finally, combining with (\ref{mabel}) and  lemma \ref{t3}:
$$
\frac{1}{n} H(\CQ^n| X^n \in \CF) = H(\rho_{\CF}) \leq H(\CQ|U) + \frac{1}{n} + 
\epsilon \log d + c \delta'.
$$
For sufficiently small $\epsilon \leq \delta'$, and setting 
$n_2 = \max \{ n_0, n_1, \delta'^{-1} \}$, 
(\ref{treca}) follows with
$$\delta' = \frac {\delta}{(2 + |\CX|)c + 1 + \log d}.$$
\qed

\vspace{2mm}

\begin{cor5} \label{t5}
For every $\epsilon, \delta > 0$ and $n \geq n_2(|\CU|,|\CX|,d,\delta,\epsilon)$
there exists a function $g: \CX^n \rightarrow \CU^n$ such that
\beq
\frac{1}{n} H(\CQ^n| g(X^n)) \leq H(\CQ|U) + \delta,
\label{ceta}
\eeq
\beq
 \left | \frac{1}{n} H(X^n| g(X^n)) - H(X|U) \right | \leq \delta.
\label{peta}
\eeq
\end{cor5}
\noindent  $\mathbf{Proof }  $ \space \space
Again it suffices to prove the claim for sufficiently small $\epsilon$.
By an iterative application of lemma  \ref{t4} we can find disjoint subsets
$\CF_1,  \dots, \CF_M$ of $\CX^n$ such that
  $$
  {\rm{Pr}}\{X^n \notin \bigcup_{\alpha = 1}^M \CF_\alpha \} \leq \epsilon
  $$
and for some sequences $u^n_\alpha \in \CT^n_{U, \delta}$, $\alpha = 1,  \dots, M$
$$
\left | \frac{1}{n} \log |\CF_\alpha|  - H(X|U) \right | \leq \frac{\delta}{2}
$$
and
$$
\frac{1}{n} H(\CQ^n| X^n \in \CF_\alpha) \leq H(\CQ|U) + \frac{\delta}{2}.
$$
Define, choosing some $u^n_0$ different from the $u^n_\alpha$,
$$
g({x^n}) =  \left\{ \begin{array}{ll}
u^n_\alpha & { {x^n} \in \CF_\alpha}\\
u^n_0 & {\rm{otherwise.}} 
\end{array} \right. 
$$
Then
$$
\frac{1}{n} H(\CQ^n| g(X^n)) \leq H(\CQ|U) + \frac{\delta}{2} + \epsilon H(\CQ)
$$
and
$$
\left | \frac{1}{n} H(X^n| g(X^n)) - H(X|U) \right | \leq 
\frac{\delta}{2} + \epsilon H(X).
$$
Finally, choose $\epsilon \leq \max \{ \frac{\delta}{2 H(\CQ)}, \frac{\delta}{2 H(X)} 
\}$. \qed

\vspace{2mm}

\noindent We are now in a position to prove the direct coding part of theorem \ref{t1}.

\vspace{1mm}

\noindent  {\bf Proof of Theorem 1 (coding) }   \space \space 
We first show that $(C,R) = (I(U;X), I(U;X) - I(U; \CQ))$ is
achievable. We  follow the classical proof \cite{ac2} closely. Define $K(X) = g(X)$.
Then $$\frac{1}{n} H(X^n|K) = H(X) - \frac{1}{n} H(K)$$  and (\ref{peta}) imply
\beq
 \left | \frac{1}{n} H(K) - I(U;X) \right | \leq \delta.
\label{sesta}
\eeq
Also by (\ref{ceta}) and (\ref{sesta}) we have
$$
\frac{1}{n} (H(K) - I(K; \CQ^n)) \leq I(U;X) - I(U; \CQ) + 2 \delta.
\label{sedma}
$$
Note that lemma \ref{t2} applied to the \emph{supersystem} $K \CQ^n$
guarantees the achievability of $(H(K), H(K) - I(K; \CQ^n)$.
Hence, for sufficiently large (super)blocklength $k$
there exists a mapping $f(K^k)$ of rate 
$\frac{1}{n k} \log |f| \leq I(U;X) - I(U; \CQ) + 2 \delta $ (here
$|f|$ is the image size of $f$), which allows $K^k$ to be reproduced with 
$\epsilon$ error. This yields an amount of $\epsilon$-randomness bounded
from below by $nk (I(U;X) - \delta)$. However, to prove the claim, we need
to show that the rate is bounded from above by exactly $I(U;X) - I(U; \CQ)$.
This is accomplished by setting the blocklength to $N = n k (1 + 2 \delta \kappa)$,
where $\kappa = \frac{1}{I(U;X) - I(U; \CQ)}$, and ignoring the last 
$ 2 \delta \kappa n k$ source outputs. Then indeed
$$ R = \frac{1}{N} \log |f| \leq I(U;X) - I(U; \CQ) $$  
while 
$$ C = \frac{1}{N} H(K^k) \geq I(U;X) - \delta(\kappa' + 2 \kappa), $$
with $\kappa' = \frac{1}{I(U;X)}$. 
\par
If now the classical communication rate $R'$ is available, we may use the procedure
outlined above to achieve a CR  rate of $I(U;X)$ while
communicating at rate $R=I(U;X)-I(U;\CQ)$, at least if $R\leq R'$.
But of course the ``surplus'' $R'-R$ is then still free to generate common
randomness trivially by Alice transmitting locally generated fair coin flips.
This shows that at communication rate $R'$, CR at rate
$$C' = R'-R + I(U;X) = R' + I(U;\CQ)$$
can be generated.
\qed

\vspace{2mm}

\noindent $\mathbf{Remark }  $ \space \space 
For $R\leq H(X)-I(X;\CQ)=H(X|\CQ)$, the maximization
constraint in (\ref{main}) may be replaced by an equality, i.e., 
\beq
D^*(R) = \tilde{D}(R)
\label{cayuga}
\eeq
where
$$
 \tilde{D}(R)= \max_{U|X} \{ I(U;\CQ)\, | \, I(U;X) - I(U;\CQ) = R \}. 
$$
To see this, note that 
$$
D^*(R) = \max_{0 \leq R' \leq R} \tilde{D}(R),  
$$
so it suffices to show that $\tilde{D}(R)$ is monotonically increasing.
This, in turn, holds if $\tilde{D}(R)$ is concave and achieves its
maximum for $R = H(X|\CQ)$. The concavity proof is virtually identical to the 
proof of lemma  \ref{t6}. The second property follows from
$$I(U;\CQ) \leq I(UX;\CQ) = I(X;\CQ)$$ 
and $I(U;X|\CQ) \leq H(X|\CQ)$. 
\par
Note that for $R \geq H(X|\CQ)$, the function $D^*(R)$ is simply
constant (and equal to $ D(\infty) = I(X;\CQ)$). 
\par\bigskip
Having established (\ref{cayuga}), we shall now relate $D^*(R)$ to
the quantum compression with classical side information trade-off curve 
$Q^*(R)$ of Hayden, Jozsa and Winter \cite{hjw}. For a classical-quantum
system $X \CQ$, given by the pure state ensemble $\{ \ket{\varphi_x}, p(x) \}$,
and $R\leq H(X)$,
$$
Q^*(R) = \min_{U|X} \{ H(\CQ|U) \, | \, I(U;X) = R \}
       = H(\CQ) - \max_{U|X} \{ I(U;\CQ) \, | \, I(U;X) = R \}.
$$
(For rates $R>H(X)$, $Q^*(R)=0$.)

The following relation to our $C^*(R)$ is now easily verified:
\beq
D^*(x) + Q^*(D^*(x) + x) = H(\CQ).
\label{keuka}
\eeq
Indeed, for $x\leq H(X|\CQ)$, and a maximizing variable $U$,
$x=I(U;X)-I(U;\CQ)$ and $D^*(x)=I(U;\CQ)$. Then, $x+D^*(x)=I(U;X)$,
so $U$ is feasible for $Q^*(x+D^*(x))$ and indeed optimal,
using once more the monotonicity of $\widetilde{D}$.

We should remark, however, that to the best of our knowledge,
eq.~(\ref{keuka}) has no simple operational meaning.
Still, it allows us to ``import'' the numerically calculated trade-off
curves from \cite{hjw} for various ensembles of interest: the curves are
then parametrized via $s=x+D^*(x)$ and $x$.

\begin{figure}
\centerline{ {\scalebox{0.50}{\includegraphics{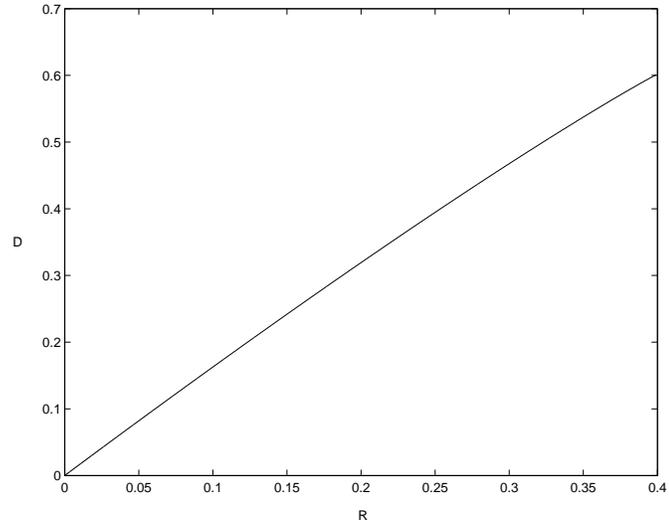}}}}
\caption{$D(R)$ for the two-state ensemble $\CE$.}
\end{figure}

Figure 1 (cf. \cite{hjw}, figure 2) shows the distillable CR-rate trade-off 
curve $D(R) = D^*(R)$ for the simple two-state ensemble 
$\CE$ given by the non-orthogonal 
pair $\{ \ket{0}, \frac{1}{\sqrt{2}}(\ket{0} +\ket{1}) \}$, 
each occurring with probability $\frac{1}{2}$.
This curve is not much better than the linear lower bound obtained
by time-sharing between $(0,0)$ and the Slepian-Wolf point 
$(1 - H(\CE), H(\CE))$, where $H(\CE)$ denotes the entropy of the average 
density matrix of the ensemble $\CE$.

\begin{figure}
\centerline{ {\scalebox{0.50}{\includegraphics{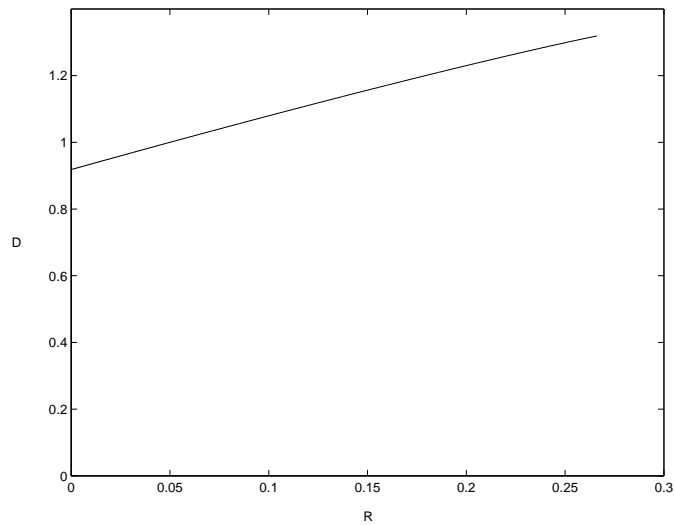}}}}
\caption{$D(R)$ for the three-state ensemble $\CE_3$.}
\end{figure}

Figure 2 (cf. \cite{hjw}, figure 4)
corresponds to the three state ensemble $\CE_3$ consisting
of the states $\ket{\varphi_1} = \ket{0}, \ket{\varphi_1} = \frac{1}{\sqrt{2}}
(\ket{0} +\ket{1})$ and $\ket{\varphi_3} = \ket{2}$ with equal probabilities.
Without any communication it is already possible to extract
$h_2(\frac{1}{3})$ bits of CR, due to Bob's ability
to perfectly distinguish whether his state is in $\{ \ket{\varphi_1}, \ket{\varphi_2} \}$
or $ \{ \ket{\varphi_3} \}$. The curve then follows a rescaled 
version of figure 1 to meet the Slepian-Wolf point 
$(H(\frac{1}{3}, \frac{1}{3}, \frac{1}{3}) - H(\CE_3),  H(\CE_3))$.

\begin{figure}
\centerline{ {\scalebox{0.50}{\includegraphics{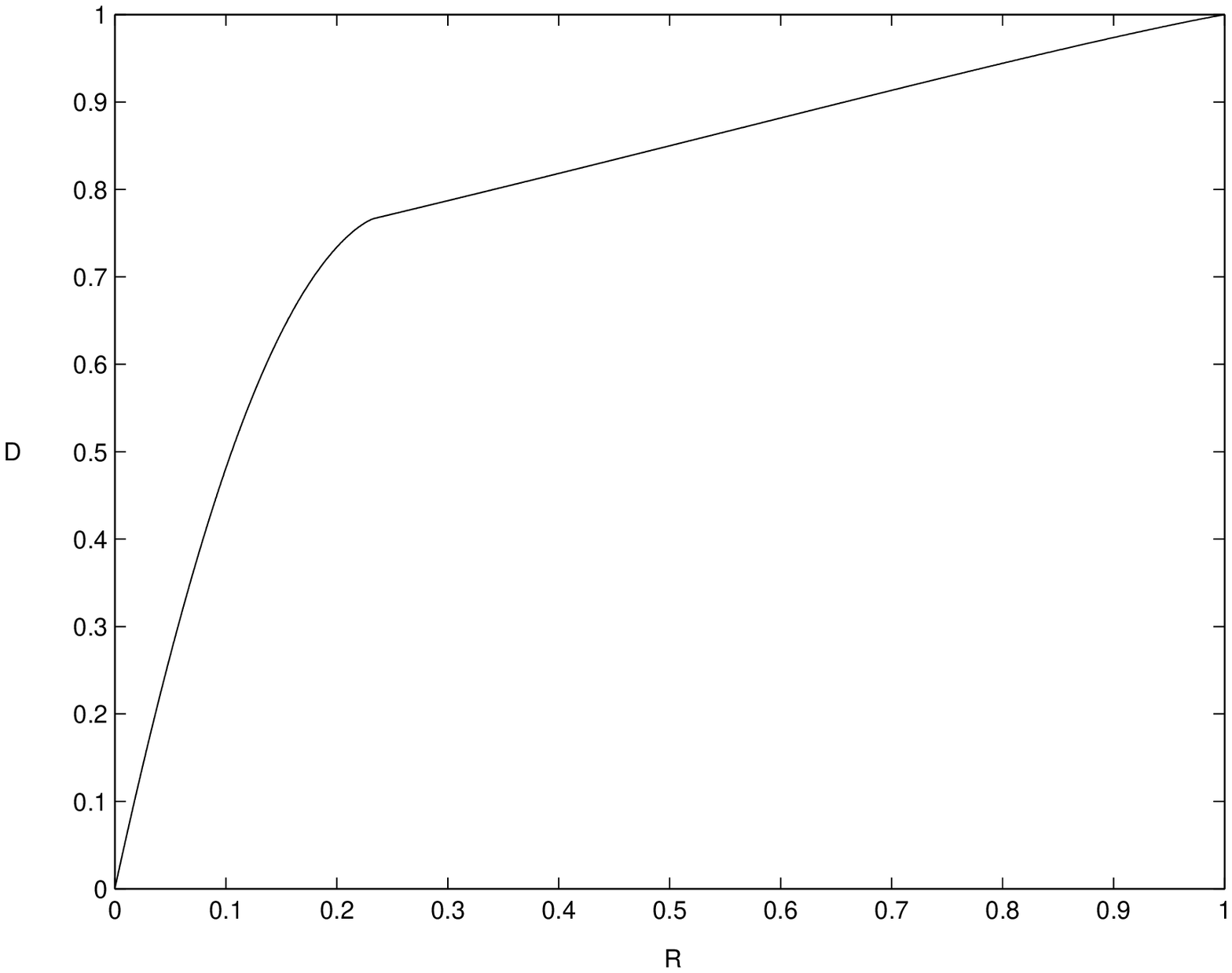}}}}
\caption{$D(R)$ for the parametrized BB84 ensemble with 
$\theta = \frac{\pi}{8}$.}
\end{figure}

Our third example is the parametrized BB84 ensemble 
$\CE_{\rm BB}(\theta)$, defined by the states
\begin{eqnarray}
\ket{\varphi_1} & = & \ket{0}   \nonumber\\
\ket{\varphi_2} & = &   \cos \theta 
\ket{0} + \sin \theta  \ket{1}  \nonumber\\
\ket{\varphi_3} & = &  \ket{1}  \nonumber\\
\ket{\varphi_4} & = & - \sin \theta \ket{0} + \cos \theta  \ket{1}   ,\nonumber
\end{eqnarray}
each chosen with probability $\frac{1}{4}$. The $D(R)$ curve for 
$\theta = \pi/8$, shown in figure 3 (cf.  \cite{hjw}, figure 5), has a special
point at which the slope is discontinuous.    
For $0 < \theta \leq \pi/4$, $\CE_{\rm BB}(\theta)$ has a natural coarse graining
to the ensemble consisting of two equiprobable mixed states, 
$\frac{1}{2}(\ket{\varphi_1}\bra{\varphi_1}) +  \ket{\varphi_2}\bra{\varphi_2})$
and
$\frac{1}{2}(\ket{\varphi_3}\bra{\varphi_3}) +  \ket{\varphi_4}\bra{\varphi_4})$.
The special point is precisely the Slepian-Wolf point for this
coarse-grained ensemble, treating $\ket{\varphi_1}$ and $\ket{\varphi_2}$,
and $\ket{\varphi_3}$ and $\ket{\varphi_4}$ as indistinguishable.

\begin{figure}
\centerline{ {\scalebox{0.50}{\includegraphics{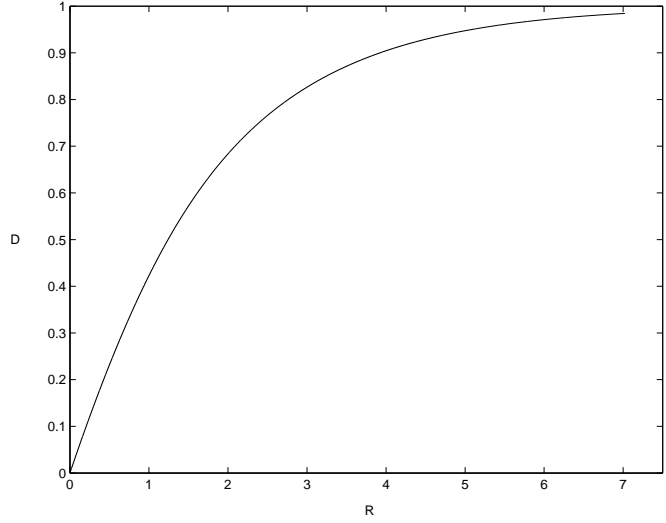}}}}
\caption{$D(R)$ for the uniform ensemble.}
\end{figure}

Finally, figure 4. (cf.  \cite{hjw}, figure 5 and \cite{devberger})
shows $D(R)$ for the uniform qubit ensemble, 
a uniform distribution of pure states over the Bloch sphere.
Strictly speaking, theorem 1 should be extended to include
continuous ensembles; we shall not do this here, but merely conjecture it
and refer the reader to  \cite{hjw} for an example of such an extension.
The curve approaches $D = 1$ only in the $R \rightarrow \infty$ limit. 
It has an explicit parametrization computed from (\ref{keuka}) and 
\cite{devberger}:
\begin{eqnarray}
R & = &  h_2 \(\frac{1}{\la} - \frac{1}{e^\la - 1} \) +
\frac{\la}{e^\la - 1} - 2 + \log \(\frac{\la e^\la}{e^{\la} - 1} \)  \nonumber\\ 
D(R) & = & 1 -  h_2 \(\frac{1}{\la} - \frac{1}{e^\la - 1} \)   \ \nonumber
\end{eqnarray}
for $\lambda \in (0, \infty)$, where $h_2(p) = - p \log p
 - (1 - p) \log (1 - p)$ is the binary Shannon entropy.

\section{General quantum correlations}
\label{general}

Consider the following double-blocking protocol for the case of 
$\{ q \, q \}$ resources: given a word of length $n L$, 
Alice performs the same measurement on each of the $n$ blocks of length $L$.
This leaves her with $n$ copies of the resulting $\{ c \, q \}$ resource, 
to which we apply the $\{ c \, q \}$ protocol described in the previous
section. Letting $n \rightarrow \infty$ and then $L \rightarrow \infty$
yields the same results as the most general protocol described in Section 1.
Let us assume $L = 1$ for the moment.   
The measurement $\CM$ on Alice's subsystem $\CA$,
defined by the positive operators $(E_x)_{ x \in \CX }$ with  $\sum_x E_x = \1$,
may be thought of as a map sending a quantum system $\CA \CB$
in the state 
$\rho^{\CA \CB}$ to a classical-quantum system $X \CQ$
in the state given by the ensemble $\{ \rho_x, p(x) \}$, where 
\begin{eqnarray}
p(x)    & = &  \tr_{\!\CA} \, \left( \rho^\CA E_x \right),  \nonumber\\
\rho_x  & = & \frac{1}{p(x)} 
        \tr_{\!\CA} \, \left ((\sqrt{E_x} \otimes \1) \rho^{\CA \CB} (\sqrt{E_x} \otimes \1) \right ).  \nonumber
\end{eqnarray}
All the relevant information is now encoded in
the shared ensemble. Theorem 1 now applies, yielding an expression
for the $L = 1$ CR-rate curve:
\beq
C^{(1)}(R) = R + \max_{\CM: {\CA \CB} \mapsto X \CQ}
               \,\, \max_{U|X} \bigl\{ I(U;\CQ) \, | \, I(U;X) - I(U; \CQ) \leq R \bigr\}. 
\label{gen1}
\eeq

Similarly we have
\beq \label{D1:infty}
D^{(1)}(\infty) = \max_{\CM: {\CA \CB} \mapsto X \CQ} I(X;\CQ),
\eeq
which is precisely the classical correlation measure 
$C_{\CA} (\rho^{\CA \CB})$ proposed in \cite{hv}. Note that
w.l.o.g.~we may assume the measurement to be rank-one,
and $|{\cal X}|\leq d^2$, $d$ the dimension of the
${\cal A}$-system, because a non-extremal POVM cannot be optimal.

However, in  general one must allow for ``entangling'' measurements performed on an
arbitrary number $L$ copies of $\rho^{\CA \CB}$, 
yielding an expression for  $C^{(L)}(R)$ analogous to
(\ref{gen1}):
$$
C^{(L)}(R) =  R + \max_{\CM: {\CA^L \CB^L} \mapsto X \CQ}
               \,\, \frac{1}{L} \max_{U|X} \bigl\{ I(U;\CQ) \, | \, I(U;X) - I(U;\CQ) \leq R \bigr\}. 
\label{gen2}
$$
Finally, taking the large $L$ limit gives
$$
 C(R) = \lim_{L \rightarrow \infty} C^{(L)}(R).
$$
Similarly
$$
 D(\infty) = \lim_{L \rightarrow \infty} D^{(L)}(\infty),
$$
which is the ``regularized'' version of 
$D^{(1)}(\infty)$ and the more appropriate asymmetric 
measure of classical correlations present in the bipartite state $\rho^{\CA \CB}$.
It is an interesting question whether $L = 1$ suffices to attain $C(R)$, 
or at least $D(\infty)$. In the remainder of this section we present
some partial results concerning this issue.

\begin{expl} \label{accessible}
Let Alice and Bob switch roles: consider a state
$$\rho^{\CA\CB} = \sum_x p(x)\rho_x^{\CA} \otimes \ket{x}\bra{x}^{\CB},$$
i.e. now Alice holds the ensemble states $\rho_x$
while Bob has the classical information $x$, with probability $p(x)$. 
\par
According to (\ref{D1:infty}), $D^{(1)}(\infty)$ is equal to the
\emph{accessible information} of the state ensemble $\CE=\{ \rho_x,p(x) \}$,
denoted $I_{\rm acc}(\CE)$~\cite{holevo2}.
On the other hand, we know from~\cite{holevo2} that
$I_{\rm acc}(\CE\otimes\CE') = I_{\rm acc}(\CE)+I_{\rm acc}(\CE')$,
for a second ensemble $\CE'$, hence
$$D(\infty) = D^{(L)}(\infty) = D^{(1)}(\infty) = I_{\rm acc}(\CE).$$
\end{expl}

This single-letterization of the accessible correlation can, in fact, 
be generalized to arbitrary separable states. Indeed, the following holds,
in some analogy to the additivity of capacity for entanglement breaking
channels~\cite{qcq} (we include the state dependence in our notation of
$D^{(1)}$ etc.):
\begin{thm1} \label{dist-add}
  Let $\rho^{AB}$ be separable and $\sigma^{A'B'}$ be arbitrary. Then,
  $$D^{(1)}(\rho\otimes\sigma,\infty) = D^{(1)}(\rho,\infty)+D^{(1)}(\sigma,\infty).$$
  From this, by iteration, we get of course
  $$D(\rho,\infty) = D^{(L)}(\rho,\infty) = D^{(1)}(\rho,\infty).$$
\end{thm1}

\noindent  $\mathbf{Proof }  $ \space \space 
  $D^{(1)}(\rho\otimes\sigma,\infty) \geq D^{(1)}(\rho,\infty)+D^{(1)}(\sigma,\infty)$
  is trivial for arbitrary states, for we can always use product measurements.
  For the opposite inequality,
  we write $\rho$ as a mixture of product states:
  $$\rho^{\CA\CB} = \sum_j q_j \hat{\tau}_j^{\CA} \otimes \tau_j^{\CB},$$
  which can be regarded as part of a classical-quantum system $J \CA \CB $
with EHS representation  
  $$\rho^{\cal J\!AB} = \sum_j q_j \ket{j}\bra{j}^{\CJ} \otimes\hat{\tau}_j^{\CA} 
\otimes \tau_j^{\CB}, $$
  whose partial trace over $\CJ$ it obviously is.
  \par
  Now we consider a measurement ${\cal M}=(E_x)_{x\in{\cal X}}$ on the combined
  system ${\cal AA'}$. Then, by definition, the post--measurement states on
  ${\cal BB'}$ and the probabilities are given by
  \begin{eqnarray}
    p(x)\rho_x &=& \tr_{\!\cal AA'} \left[ \bigl(\rho^{\cal AB}\otimes\sigma^{\cal A'B'}\bigr)
                                              \bigl(E_x^{\cal AA'}\otimes\1\bigr) \right] \nonumber\\
               &=& \sum_j q_j \tau_j^{\CB} \otimes
                       \tr_{\!\cal AA'}\left[ \bigl(\hat{\tau}_j^{\CA}\otimes\sigma^{\cal A'B'}\bigr)
                                                             \bigl(E_x^{\CA \CA'}\otimes\1\bigr) \right] 
                                                                                          \nonumber\\
               &=& \sum_j q_j \tau_j^{\CB} \otimes
                       \tr_{\!\CA'}\left[ \sigma^{\cal A'B'}\bigl(F_{x|j}\otimes\1\bigr) \right],
                                                                                          \nonumber
  \end{eqnarray}
  with the POVMs ${\cal N}_j=(F_{x|j})_{x\in{\cal X}}$ on ${\cal A'}$,
  labeled by the different $j$:
  $$F_{x|j} = \tr_{\!\CA}\bigl( E_x(\hat{\tau}_j\otimes\1) \bigr).$$
  Thus, applying the measurement $ \CM$ on $\CA \CA'$ on
  $\rho^{\cal J\!AB}\otimes\sigma^{\cal A'B'}$, 
  and storing the result in $X$ leads to the classical-quantum system
  $XJ \CB \CB'$ defined by the EHS state
  $$\omega = \sum_{x,j} \ket{x}\bra{x}^{\CC} \otimes q_j \ket{j}\bra{j}^{\CJ} \otimes
                         \tau_j^{\CB} \otimes
                         \tr_{\!\CA'}\left[ \sigma^{\cal A'B'}\bigl(F_{x|j}\otimes\1\bigr) \right].$$
  With respect to it,
\begin{eqnarray}
I(X;\CB \CB') & =  &  I(X; \CB) + I(X; \CB'| \CB)   \nonumber \\
     &= &  I(X;\CB) + I(X \CB;\CB')-I(\CB;\CB')                     \nonumber \\
   &   = &   I(X;\CB)+I(X\CB;\CB')             \nonumber \\
     & \leq  & I(X;\CB)+I(XJ;\CB')             \nonumber \\
    &  =  &  I(X;\CB)+I(XJ;\CB')-I(J;\CB')       \nonumber \\
     & =  &  I(X;\CB)+I(X;\CB'|J),      \label{eq:thatsit}
  \end{eqnarray}
 using the chain rule, the fact that $\CB\CB'$ is in a product state,
 the data processing inequality \cite{ahlswede:loeber}, the fact that $J\CB'$ is in a
 product state and the chain rule once more.
 
  In (\ref{eq:thatsit}) notice that the first mutual
  information, $I(X;{\cal B})$, relates to applying the POVM ${\cal M}$ to ${\cal A}$,
  with an ancilla ${\cal A'}$ in the state $\sigma^{\CA'}$ -- but this
  can be described by a POVM ${\cal N}$ on ${\cal A}$ alone. The second,
  $I(X;{\cal B'}|J)$, is a probability average over mutual informations relating to
  different POVMs on ${\cal A'}$. Thus
  $$I(X;{\cal BB'}) \leq D^{(1)}(\rho,\infty)+D^{(1)}(\sigma,\infty),$$
  which yields the claim, as ${\cal M}$ was arbitrary.
\qed

\vspace{2mm}


\begin{expl}  \label{entangled}
For a pure entangled state $\psi=\ket{\psi}\bra{\psi}$, we can easily see that
$$D^{(1)}(\psi,\infty) =  D^{(1)}(\psi, 0) = E(\ket{\psi}) = H(\tr_{\!\CB}\psi).$$
Indeed, the right hand side is attained for Alice and Bob both measuring
in bases corresponding to a Schmidt decomposition of $\ket{\psi}$.
On the other hand, in the definition of $D^{(1)}$, eq. (\ref{D1:infty}),
the mutual information $I(X;\CQ)$ is upper bounded by $H(\CQ)$,
which is the right hand side in the above equation.

Thus, if both $\psi$ and $\varphi$ are pure entangled states,
$$D^{(1)}(\psi\otimes\varphi,\infty) = D^{(1)}(\psi,\infty)+D^{(1)}(\varphi,\infty).$$
In particular,
$$D(\psi,\infty)  = D^{(L)}(\psi,\infty) = D^{(1)}(\psi,\infty).$$
\end{expl}

More generally, we have (compare to the additivity of channel capacity
if one of the channels is noiseless~\cite{schu-west}):

\begin{thm1}  \label{ent-add}
  Let $\rho^{\CA\CB}=\ket{\psi}\bra{\psi}$ be pure and $\sigma^{\CA'\CB'}$
  arbitrary. Then
  $$D^{(1)}(\rho\otimes\sigma,\infty) = D^{(1)}(\rho,\infty) +
                                         D^{(1)}(\sigma,\infty).$$
\end{thm1}
\noindent  $\mathbf{Proof }  $ \space \space
As usual, only ``$\leq$'' has to be proved. Given any POVM
${\cal M}=(E_x)_{x \in \CX}$ on $\CA\CA'$, the classical-quantum 
correlations $X \CB \CB'$ remaining after this measurement
is performed are described by 
$$\omega = \sum_x \ket{x}\bra{x}^{\CC}\otimes
                  \tr_{\!\cal AA'}\left[ \bigl(\rho^{\cal AB}\otimes
                                              \sigma^{\cal A'B'}\bigr)
                                       \bigl(E_x^{\cal AA'}\otimes\1\bigr)
                                 \right].$$
We shall assume that $\ket{\psi}$ is in Schmidt form:
$$\ket{\psi} = \sum_j \sqrt{\lambda_j}\ket{j}^{\CA}\ket{j}^{\CB}.$$
Measuring in the basis $\ket{j}$ on $\CB$ and recording the result in
orthogonal
states $\ket{j}\bra{j}$ in a register $\CJ$ transforms $\omega$ into the
state
$$\omega' = \sum_{x,j} \lambda_j\ket{j}\bra{j}^{\CJ} \otimes
                       \ket{x}\bra{x}^{\CC} \otimes
                       \tr_{\!\cal AA'}\left[
                            \bigl(\ket{j}\bra{j}^{\CA}\otimes
                                           \sigma^{\cal A'B'}\bigr)
                            \bigl(E_x^{\cal AA'}\otimes
                                            \1^{\CB'}\bigr) \right].$$
We claim that
\beq \label{subadd-ent}
  I_\omega(X;{\cal BB'}) \leq I_{\omega'}(X;\CB'|J)+ H_\omega(\CB),
\eeq
where the subscript indicates the state relative to which the respective
information quantity is understood. Clearly, from this the theorem
follows: on the right hand side, the entropy is the entropy of
entanglement of $\rho$, and the mutual information is an average of mutual
informations for measurements $\CM_j$ on $\CA'$, defined as performing
$\CM$ with ancillary state $\ket{j}\bra{j}$ on $\CA$.
\par
To prove (\ref{subadd-ent}), we first reformulate it such that all
entropies
refer to the same state. For this, observe that the measurement
of $j$ can be done by adjoining the register $\CJ$ in a null state
$\ket{0}$, applying a unitary which maps $\ket{j}^{\CB}\ket{0}^{\CJ}$ to
$\ket{j}^{\CB}\ket{j}^{\CJ}$, and tracing out $\CB$. Denote by $\Omega$
the state obtained from $\omega$ by this procedure. Obviously then,
(\ref{subadd-ent}) is equivalent to
\beq \label{subadd-alt}
  I(X;{\CB \CJ \CB'}) \leq I(X;\CB'|\CJ)+H(\CB\CJ),
\eeq
with respect to  $\Omega$,
because isometries do not alter entropies.
\par
Now, writing out the above quantities as sums and differences of
entropies, and using the fact that $\CB\CJ-\CB'$ is in a product state, a
number of terms cancel out, and (\ref{subadd-alt}) becomes equivalent to
$$H(\CB \CJ \CB'|X) \geq H(\CB'|X\CJ).$$
But now rewriting the left hand side, using $H(\CB \CJ|X)\geq 0$ (because it
is an average of von Neumann entropies), we estimate:
\begin{eqnarray}
  H(\CB \CJ \CB'|X) &=   & H(\CB'|\CB \CJ X) + H(\CB \CJ|X) \nonumber\\
                  &\geq& H(\CB'|\CB \CJ X)              \nonumber\\
                  &\geq& H(\CB'|\CJ X),                 \nonumber
\end{eqnarray}
where in the last line we have used strong subadditivity, and we are done.
\qed

\vspace{3mm}

We do not know if additivity as in the above cases holds universally, but
we regard our results as evidence in favor of this conjecture.

Returning to finite side-communication, it is a most interesting question whether
a similar single-letterization can be performed.
We do not know if an additivity-formula, similar to the one in lemma \ref{t7}
for classical-quantum correlation, holds for the rate function
$D^{(1)}(\rho\otimes\sigma,R)$. In fact, this seems unlikely because
its definition does not even allow one to see that it is concave in $R$
(which it better had to if it be equal to the regularized quantity.).
Of course this can easily be remedied by going to the concave hull
$\widetilde{D}^{(1)}$ of $D^{(1)}$: note that both regularize to the same
function for $L\rightarrow\infty$.
However, we were still unable to prove additivity for $\widetilde{D}^{(1)}$.
This would be a most desirable property, as it
would allow single-letterization of the rate function just as in the
case of classical-quantum correlations.
As it stands, $\widetilde{D}^{(1)}(\rho,R)$ is the CR obtainable
from $\rho$ in excess over $R$, if (one-way) side communication is limited to $R$ and
\emph{if the initial measurement is a tensor product}.

\section{Discussion}
We have introduced the task of distilling common randomness from a quantum state
by limited classical one-way communication, placing it in the context of general
resource conversion problems from classical and quantum information theory.
Our exposition can be read as a systematic objective for 
the field of quantum information theory: to study all the conceivable
inter-conversion problems between the resources enumerated in the Introduction.
\par
Our main result is the characterization of the optimal asymptotically distillable
common randomness $C$ (as a function of the communication bound $R$);
in the case of initial classical-quantum correlations this characterization is
a single-letter optimization. 
\par
A particularly interesting figure is the total ``distillable common randomness'',
which is the supremum of $C(R)-R$ as $R\rightarrow\infty$: for the classical-quantum
correlations it turns out to be simply the quantum mutual information, and in general
it is identical to the regularized version of the measure for classical 
correlation put forward by Henderson and Vedral~\cite{hv}.
\par
It should be noted that this quantity is generally smaller than the
quantum mutual information $I(\CA;\CB)$ of the state $\rho^{\CA\CB}$
(which was discussed in~\cite{cerf:adami}), but larger than the quantity
proposed by Levitin~\cite{levitin}. Interestingly, while the former work
simply examines a quantity defined in formal analogy to classical mutual
information for its usefulness to (at least, qualitatively) describe quantum
phenomena, the latter motivates the definition by recurring to
operational arguments. Of course, all this shows is  that there can be
several operational approaches to the same intuitive concept: quantities
thus defined might coincide for classical systems but differ in the quantum
version.
\par
This is what we see even within the realm of our definitions. In the
classical theory~\cite{ac2} the total distillable CR equals
the mutual information of the initial distribution, regardless of the particulars
of the noiseless side communication: whether it is one-way from Alice to Bob or
vice versa, or actually bidirectional, the answer is the mutual information.
There are simple examples of quantum states where the total distillable common
randomness depends on the communication model: the classical-quantum correlation
associated with an ensemble $\CE=\{ \rho_x,p(x) \}$ of states at Bob's side
(compare eq.~(\ref{cq})) leads to $I(\CA;\CQ)=\chi(\CE)$  if one-way communication from
Alice to Bob is available. If only one-way communication from Bob to Alice is available,
it is only $I_{\rm acc}(\CE)$, the accessible information of the ensemble $\CE$,
which usually is strictly smaller than the Holevo information
$\chi(\CE)$~\cite{holevo}.
\par
An open problem left in this work is to decide the additivity
questions in section \ref{general}: is the distillable common randomness
$D^{(1)}(\rho,\infty)$ additive in general? Does the rate function
$D^{(1)}(\rho,R)$ obey an additivity-formula like the one in lemma \ref{t7}?
Finally, there is the issue  of finding the ``ultimate'' 
distillable common randomness involving two-way communication.

\bigskip

\noindent {\bf{Acknowledgments}} \, 
We thank C. H. Bennett, D. P. DiVincenzo, B. M. Terhal, 
J. A. Smolin and R. Abbot for useful discussions. ID's work was supported
in part by the NSA under the US Army Research Office (ARO), 
grant numbers DAAG55-98-C-0041 and DAAD19-01-1-06. 
AW is supported by the U.K.~Engineering and Physical Sciences
Research Council.

\appendix

\section{Appendix}

We shall list definitions and properties of
typical sequences and subspaces \cite{ck, nono, winter}. Consider
the classical-quantum system $U X \CQ$ in the state
defined by the ensemble $\{ p(u,x), \rho_{ux} \}$. 
$X$ is defined on the set $\CX$ of cardinality $s_1$  
and $U$ on the set $\CU$ of cardinality $s_2$.
Denote by $p(x)$ and $P(x|u)$ the distribution of $X$ 
and conditional distribution of $X|U$ respectively.

For the probability distribution $p$ on the set $\CX$
define the set of \emph{typical sequences} (with $\delta>0$)
$$\CT^n_{p,\delta}=\left\{x^n:\forall x\ | N(x|x^n)- n p(x)|\leq
                              n {\delta} \right\},$$
where $N(x|x^n)$ counts the number of occurrences of
$x$ in the word $x^n=x_1\ldots x_n$ of length $n$.
When the distribution $p$ is associated with some random variable $X$
we may use the notation $\CT^n_{X,\delta}$. 

For the stochastic matrix $P: \CU \rightarrow \CX$ and
$u^n \in  \CU^n$ define the set of \emph{conditionally typical sequences} 
(with $\delta>0$) by
$$\CT^n_{P,\delta}(u^n) = \left\{x^n:\forall u,x \ | N((u,x)| (u^n,x^n)) 
- P(x|u) N(u|u^n)| \leq n{\delta} \right\}.$$
When the stochastic matrix $P$ is associated with some conditional 
random variable $X|U$  we may use the notation 
$\CT^n_{X|U,\delta}(u^n)$.

For a density operator $\rho$ on a $d$-dimensional Hilbert space $\CH$, 
with eigen-decomposition $\rho = \sum_{k = 1}^{d} \lambda_k \ket{k}\bra{k}$ 
define (for $\delta>0$) the \emph{typical projector} as
$$
\Pi^n_{\rho,\delta}=\sum_{k^n\in{\CT}^n_{R,\delta}}
\ket{k^n}\bra{k^n}.
$$
When the density operator $\rho$ is associated with some quantum  
system $\CQ$ we may use the notation 
$\Pi^n_{\CQ,\delta}$.

For a collection of states ${\rho}_u$, $u \in \CU$, and
$u^n\in \CU^n$ define the \emph{conditionally typical projector} as
$$\Pi^n_{\{\rho_u\},\delta}(u^n)=\bigotimes_u 
                              \Pi^{I_u}_{\rho_u,\delta},$$
where $I_u=\{i:u_i=u\}$ and $\Pi^{I_u}_{\rho_u,\delta}$
denotes the typical projector of the density operator ${\rho}_u$
in the positions given by the set $I_u$ in the tensor product of $n$ factors.
When the $\{\rho_u \}$ are associated with some conditional classical-quantum system  
system $\CQ|U$ we may use the notation 
$\Pi^n_{\CQ|U,\delta}(u^n)$.
We shall give several known properties of these projectors,
some of which are used in the main part of the paper. 
For any positive $\epsilon, \delta$ and $\delta'$, 
some constant $c$ depending on the particular ensemble of $UX \CQ$,
and for sufficiently large $n \geq n_0(\epsilon, \delta ,\delta')$,
the following hold.
Concerning the quantum system $\CQ$ alone:
 
\begin{eqnarray}
\tr \Pi^n_{\CQ,\delta}  & \leq &  2^{n (H(\CQ) + c \delta)} \nonumber\\
\tr \rho^{\otimes n} \Pi^n_{\CQ,\delta}  & \geq &  1 - \epsilon. \nonumber
\end{eqnarray}

Concerning the classical-quantum system $X \CQ$, and for $x^n \in \CT^n_{X,\delta'}$:

\begin{eqnarray}
\tr \Pi^n_{\CQ|X,\delta}(x^n)  &  \leq  & 2^{n (H(\CQ|X) 
+ c (\delta + \delta'))} \label{hu} \\
\tr \rho_{x^n} \Pi^n_{\CQ|X,\delta}(x^n)  & \geq &  1 - \epsilon \nonumber\\
\tr \rho_{x^n} \Pi^n_{\CQ, \, \delta + |\CX| \delta'} & \geq &  1 - \epsilon. \label{hu2}
\end{eqnarray}

These have been proven in \cite{strong}. 
Finally, concerning the full  classical-quantum system $U X \CQ$,
for $x^n \in \CT^n_{X|U,\delta'}(u^n)$
(\ref{hu2}) easily extends to
\beq
\tr \rho_{u^n x^n} \Pi^n_{\CQ|U, \, \delta + |\CX| \delta'}  \geq  1 - \epsilon. 
\label{hu3}
\eeq

\end{document}